     \documentstyle[12pt]{article}

     \catcode`\@=11

     \@addtoreset{equation}{section}

\begin{document} \baselineskip 24pt
\newcommand{\numero}{SHEP 92/93--23} 
number \newcommand{\titre}{MAJORANA NEUTRINO MASSES CAN
SAVE} \newcommand{\titreb}{ONE FAMILY TECHNICOLOUR MODELS}
\newcommand{\auteura}{N.J. Evans} \newcommand{\addressa}{ }
\newcommand{\auteurc}{D.A. Ross }
\newcommand{\beq}{\begin{equation}}
\newcommand{\eeq}{\end{equation}}
\newcommand{\Fn}{\mbox{$F(p^2,\Sigma)$}}

     \newcommand{\addressc}{Physics Department \\ University
of Southampton\\ Southampton SO9 5NH. \\ U.K. }
\newcommand{\abstrait}{ We make non perturbative estimates
of the electroweak radiative correction parameter $S$ in
dynamical symmetry breaking models with Majorana neutrino
masses. The Majorana masses are treated as perturbations to
a Non Local Chiral Model of the strong interactions. We
argue that parameter ranges exist that would allow realistic
values of $S$ and $T$ in one family Technicolour models}
\begin{titlepage} \hfill \numero \vspace{.5in}
\begin{center} {\large{\bf \titre }} {\large{\bf \titreb}}
\bigskip \\by\bigskip\\ \auteura \\ and \\ \auteurc \bigskip
\\ \addressc \\

     \renewcommand{\thefootnote}{ } \vspace{.9 in} {\bf
Abstract} \end{center} \abstrait \bigskip \\ \end{titlepage}

     \def\id{\rlap{1}\hspace{0.15em}1}

     \section{Introduction}

     Precision measurements of electroweak radiative
corrections have begun to constrain models of electroweak
symmetry breaking (EWSB).  Any new particles at the $TeV$
scale that transform under $SU(2)_L$ do not decouple from
the theory at low energies \cite{lynn} but contribute to
precision electroweak measurements through gauge boson self
energies (oblique corrections).  When the new particles have
masses much larger than the Z boson mass, $M_Z$, their major
contribution to the oblique corrections can be described by
the parameters $S$ and $T$ introduced by Peskin and Takeuchi
\cite{peskin}

     \beq S \hspace{.5cm} = \hspace{.5cm} 16 \pi
\frac{d}{dq^2} \left[ \Pi_{33}\left(q^2\right) - \Pi_{3Q}
\left( q^2 \right)\right] \mid _{q^2=0} \hspace{.5cm} =
\hspace{.5cm} - 8 \pi \frac{d}{dq^2} \left[ \Pi_{3Y}
\left(q^2\right) \right] \mid _{q^2=0} \eeq \beq \alpha T
\hspace{.5cm} = \hspace{.5cm} \delta \rho_* \hspace{.5cm} =
\hspace{.5cm} \frac{e^2}{s^2M_W^2}\left[ \Pi_{11}(0) -
\Pi_{33}(0)\right] \eeq

     $M_W$ is the W  boson mass,  $e$ is the electromagnetic
charge    and   $s^2    \equiv    \sin^2(\theta_w)$.     The
$\Pi_{ij}(q^2)$  are the gauge boson  self energies with the
indices $i,j=1,2,3,Q$  and  $Y$ referring to  the $SU(2)_L$,
electromagnetic  and  hypercharge currents respectively. The
contribution  to the self energies  from  loops of  standard
model particles are removed to isolate the dependence on the
new heavy  particles. The parameter $T$ is directly  related
to the $\rho$ parameter by $\alpha T = \delta \rho_* =
\frac{M_W^2}{M_Z^2} - 1$. Recent measurements
\cite{peskin,measure} suggest that $S \leq 0.6$ and $T\leq1
(\delta \rho_* \leq 0.7\%)$ at the $95\%$ confidence level
(c.l.)  with slightly negative values ($S$ and $T\approx
-0.5$) favoured.

     In models such as Technicolour (TC) \cite{TC} in which
EWS is broken dynamically new fermions are strongly
interacting. Estimates have been made of these new fermions'
contribution to $S$ and $T$ in the isospin preserving limit
by ``scaling up" QCD \cite{peskin} and in a non- local
chiral model of the strong interactions \cite{holdom,tern}.
These estimates suggest that the contributions to $S$ and
$T$ per fermion doublet are

     \beq \Delta S \simeq 0.1 N_{TC} \label{eq:naives},
\hspace{1cm} \Delta T = 0 \eeq

     \noindent where $N_{TC}$ is the number of
Technicolours.  All but the most minimal TC models appear to
be ruled out by these estimates of the $S$ parameter.

     However, a complete model of EWSB must account for all
the fermion masses in the Standard Model.  In particular it
must explain why these masses break isospin symmetry $eg$
$m_t/m_b \geq 20$ and why the neutrinos are massless or
nearly so.  We might expect that in a realistic TC model
isospin would be broken in the techni-fermion sector. In
this paper we investigate TC models in which the isospin
symmetry is broken by techni-neutrino Majorana masses.
Allowing Majorana masses into the theory restricts the TC
group to those with real representations.

     In one loop perturbation theory it has been shown that
Majorana masses can give negative contributions to both $S$
\cite{gatesandtern} and $T$ \cite{bert}.  The
techni-fermions, however, are strongly interacting and hence
a non-perturbative calculation must be performed to estimate
their contributions to $S$ and $T$.  The contribution to $T$
from Majorana techni-neutrinos has been estimated
\cite{majT} in Dynamical Perturbation Theory \cite{DPT} and
show the same qualitative behaviour as the perturbative
results.  The most stringent constraint on TC models comes
though from the $S$ parameter. We estimate the contributions
to $S$ by perturbing the isospin symmetry in the Non Local
Chiral Model \cite{holdom,tern}.  We argue that parameter
ranges exist that would allow realistic values of $S$ and
$T$ in one family TC models.

     In Section 2 of this paper we will review the isospin
preserving Non Local Chiral Model and the derivation of the
estimate of $S$ in Ref \cite{tern} (Eq(\ref{eq:naives})). In
Section 3 we generalize the model to include Majorana
neutrino masses. In Section 4 we present our results for the
contribution to $S$ from Majorana techni-neutrinos and
present a number of realistic techni-fermion mass spectra in
one family TC models. Section 5 concludes the paper.

     \section{The Non Local Chiral Model}

     In this section we review the derivation of
Eq(\ref{eq:naives}) in the Non Local Chiral Model
\cite{holdom,tern} approximation to TC dynamics. Consider a
techni-fermion doublet, $\Psi = \left( \begin{array}{c} U \\
D \end{array} \right)$, which transforms under some isospin
preserving TC group. The techni-fermions also transform
under the Standard Model electroweak symmetry group $SU(2)_L
\otimes U(1)_Y$.  The TC group becomes strongly interacting
at some scale $\Lambda_{TC}$ and breaks the doublets chiral
symmetry

     \beq SU(2)_L \otimes SU(2)_R \rightarrow SU(2)_V
\label{eq:symb} \eeq

     The techni-fermions acquire a dynamical self energy
term, $\Sigma(k^2)$, which acts as the order parameter for
chiral symmetry breakdown. There are three Goldstone bosons,
$\pi^a$ ($a = 1..3$), associated with the three broken
generators which are bound states of the techni-quarks and
correspond to the pions of QCD. Below the scale
$\Lambda_{TC}$ the effective theory of the Goldstone bosons
and $SU(2)_L \otimes U(1)_Y$ gauge bosons can be described
by a gauged chiral Lagrangian \cite{georgibook}.  Following
the usual formalism we write the Goldstone fields as

     \beq U = exp(i \pi^a \lambda_a / f_{\pi})   \eeq

     \noindent where $\lambda_a$ are the broken, axial
generators and $f_{\pi}$ is a dimensionful constant $\simeq
\Lambda_{TC}$.  Under global $SU(2)_L \otimes SU(2)_R$
symmetry transformations $U$ transforms as

     \beq U \longrightarrow R^{\dagger} U L \eeq

     At low energies we may perform an expansion in powers
of momentum and in the notation of Gasser and Leutwyler
\cite{gasser} the chiral Lagrangian is given up to fourth
order in the (covariant) derivative of $U$ by

     \newpage

     \begin{eqnarray} {\cal L} & = & \frac{f_{\pi}^2}{4} tr
(D^{\mu} U D_{\mu} U^{\dagger}) + L_1 tr(D^{\mu} U D_{\mu}
U^{\dagger})^2 + L_2 tr((D^{\mu} U D^{\nu}
U^{\dagger})(D_{\mu} U D_{\nu} U^{\dagger}) \nonumber \\ & &
+ L_3 tr(D^{\mu} U D_{\mu} U^{\dagger} D^{\nu} U D_{\nu}
U^{\dagger}) + iL_9 tr(F^R_{\mu \nu} D^{\mu} U D^{\nu} +
F^L_{\mu \nu}D^{\mu} U D^{\nu}) \\ & & + L_{10}
tr(U^{\dagger} F^R_{\mu \nu} U F^{L \mu \nu}) + L_{11}
tr(D^2 U D^2 U^{\dagger}) \label{eq:chirL} \nonumber
\end{eqnarray}

     The only contribution to the gauge boson self energy
$\Pi_{3Y}$ at order $q^2$ is given by $L_{10}$ \cite{l10=s}
and hence the electroweak precision parameter $S$ is given
by

     \beq S = -16 \pi L_{10} \label{eq:l10=s} \eeq

     The parameter $L_{10}$ is not determined by the chiral
symmetry and must be estimated in some approximation to the
full TC dynamics. The Non Local Chiral Model is such an
approximation. The model assumes that the following elements
of the TC dynamics are responsible for the non perturbative
features of TC above the symmetry breaking scale,
$\Lambda_{TC}$

     \noindent $\bullet$ The theory is chiral {\it ie} it
must possess an $SU(2)_L \otimes SU(2)_R$ symmetry.

     \noindent $\bullet$ The techni-fermions, $\Psi$,
acquire a dynamical self energy term, $\Sigma(k^2)$.

     \noindent $\bullet$ There are three Goldstone bosons
associated with the three generators broken in the chiral
\mbox{\hspace{.2cm} symmetry} break down.

     The Goldstone bosons appear in the theory, just as in
the Higgs model, as fluctuations about the order parameter
of symmetry break down which here is $\Sigma(k^2)$.  These
approximations are born out by the success of the model at
reproducing the low energy predictions of QCD
\cite{holdomQCD}. The Lagrangian for the techni-fermions is
thus given by

     \beq {\cal L} = \bar{\Psi}(x) \delta(x-y) \slash
\!\!\!\partial \Psi(y) + \bar{\Psi}(x) \Sigma_{\pi}(x,y)
\Psi(y) \eeq

     \noindent where the form of $\Sigma_{\pi}(x,y)$ is
determined by the chiral symmetry \cite{holdom} to be

     \begin{eqnarray} \Sigma_{\pi}(x,y) & = & \Sigma(x-y) [
1 - \frac{i}{f_{\pi}} \gamma_5[\pi(x) + \pi(y)] \nonumber \\
& & - \frac{1}{2} f_{\pi}^2[ \pi^2(x) + \pi^2(y)
+\pi(x)\pi(y) + \pi(y)\pi(x)] + ....]  \end{eqnarray}

     The action for this Lagrangian can be made locally
gauge invariant by the insertion of one or more path ordered
exponential of a line integral over the gauge field

     \beq exp \left[ -ieQ \int^y_x A_{\nu} dw^{\nu} \right]
\eeq

     \noindent where $Q$ is the charge matrix for the
fermions $\Psi$. This procedure is followed explicitly in
Ref\cite{tern} and the Feynman rules in Appendix 1 are
obtained.

     The Non Local Chiral Model has the same chiral symmetry
as the Chiral Lagrangian, Eq(\ref{eq:chirL}).  Thus the
Chiral Lagrangian may be obtained simply by integrating out
the techni-quarks from the non-local theory leaving the
Goldstone bosons and gauge bosons.  We wish to find the
coefficient $L_{10}$ in the Chiral Lagrangian; we note that
only the term in Eq(\ref{eq:chirL}) with coefficient
$L_{10}$ gives a contribution to the scattering between two
vector gauge bosons and two Goldstone bosons of the form
$q^2 g^{\mu \nu}$ where $q$ is the incoming gauge boson
momentum. $L_{10}$ is therefore given by the coefficient,
$C$, of the $q^2 g_{\mu \nu}$ term in the Taylor expansion
of the amplitude given by the sum of diagrams in Fig 1.

     \beq \begin{array}{r c r} L_{10} [
tr(2\lambda^c\lambda^a\lambda^d\lambda^b +
2\lambda^c\lambda^b\lambda^d\lambda^a) & = &
\frac{f_{\pi}^2}{8} C
[tr(2\lambda^c\lambda^a\lambda^d\lambda^b +
2\lambda^c\lambda^b\lambda^d\lambda^a) \\ & & \\
-tr(\lambda^c\lambda^d\lambda^a\lambda^b -
\lambda^c\lambda^a\lambda^b\lambda^d & &
-tr(\lambda^c\lambda^d\lambda^a\lambda^b -
\lambda^c\lambda^a\lambda^b\lambda^d \\ & & \\ -
\lambda^c\lambda^d\lambda^b\lambda^a -
\lambda^c\lambda^b\lambda^a\lambda^d)] & & -
\lambda^c\lambda^d\lambda^b\lambda^a -
\lambda^c\lambda^b\lambda^a\lambda^d)] \end{array} \eeq

     The calculation is simplified by taking the limit in
which the Goldstone bosons' momenta vanish.  The
coefficient, $C$, is an integral expression containing over
$70$ terms involving $\Sigma(k^2)$ and its derivatives which
we will not reproduce here (see Refs \cite{ternphd},
\cite{myphd}). We note that the two group theory factors
provide independent checks on $C$.  We will show in Section
4 that for a suitable choice of function for $\Sigma(k^2)$
this estimate for $S$ gives the result in
Eq(\ref{eq:naives}).

     \section{Majorana Techni-Neutrinos}

     We now consider perturbing the Non Local Chiral Model
of a techni-lepton doublet, $\psi = \left( \begin{array}{c}
N \\ E \end{array} \right)$, by explicitly introducing a
Majorana mass term, $M$, for the right handed
techni-neutrino, $N_R$. We shall assume that $M$ is small
relative to the fermions' Dirac self energy, $\Sigma(k^2)$,
so that the chiral symmetry breaking pattern
Eq(\ref{eq:symb}) is only slightly perturbed. In practice
perturbing the effective Chiral Lagrangian of QCD by
introducing current masses for the quarks provides a good
description of QCD's low energy interactions even for the
strange quark which has a current mass $\simeq
\Lambda_{QCD}$ (see for example \cite{georgibook}). Whereas
the introduction of a Majorana mass for $N_R$ explicitly
breaks the custodial $SU(2)$ symmetry and therefore alters
the symmetry breaking pattern from Eq(\ref{eq:symb}) it is
reasonable to assume that Eq(\ref{eq:symb}) is a good
approximation for Majorana masses $\leq {\cal
O}(\Lambda_{TC})$. We quantify the accuracy of this
assumption in Section 4 by comparing our results for $S$
from a lepton doublet with hard mass terms with the
perturbative one loop results.

     Writing the left and right handed degrees of freedom of
N as two Majorana (self conjugate, $\psi^M = C(\bar
\psi^M)^T$) fields $N_1^0 = \left( \begin{array}{c} N_L \\
N_L^C \end{array} \right)$ and $N_2^0 = \left(
\begin{array}{c} N^C_R \\ N_R \end{array} \right)$ the mass
terms are

     \beq {\cal L}_M = - \bar E \Sigma_E E - \frac{1}{2}
          ( \bar N_1^0 \bar N_2^{0})
          \left( \begin{array}{cc}
          0 & \Sigma_N \\
          \Sigma_N & -M
          \end{array} \right)
          \left( \begin{array}{c} N_1^{0} \\ N_2^0
     \end{array} \right) \eeq

     \noindent where $\Sigma_E$ and $\Sigma_N$ are the self
energies of the E and N fields after chiral symmetry
breakdown. The Majorana mass eigenstate fields $N_1$ and
$N_2$ with masses $M_1(p)$ and $M_2(p)$ are given, using the
notation of Ref\cite{gatesandtern}, as

     \beq \left(\begin{array}{c} N_1^0\\ N_2^{0} \end{array}
\right) = \left( \begin{array}{cc} ic_{\theta} & s_{\theta}
\\ -is_{\theta} & c_{\theta} \end{array} \right) \left(
\begin{array}{c} N_1 \\ N_2 \end{array} \right) \eeq

     \noindent where

     \beq c^2_{\theta} = \frac{M_2}{M_1+M_2}, \hspace{1cm}
s^2_{\theta} = \frac{M_1}{M_1+M_2} \eeq

     \noindent and

     \beq M = M_2 - M_1, \hspace{1cm}
     \Sigma_N = \sqrt {M_1 M_2} \eeq

     The currents that give the verticies in Appendix 1 must
be written in terms of the mass eigenstate fields, $N_1$ and
$N_2$, according to

     \beq \begin{array}{lcl} \bar{\Psi} \gamma^{\mu} \Psi &
= & \frac{1}{2}(s_{\theta}^2-c_{\theta}^2) (\bar{N_1}
\gamma^{\mu} \gamma_5 N_1 - \bar{N_2} \gamma^{\mu} \gamma_5
N_2)\\ & & + i s_{\theta} c_{\theta} (\bar{N_1} \gamma^{\mu}
N_2 - \bar{N_2} \gamma^{\mu} N_1) \\ & & \\
          \bar{\Psi} \gamma_5 {\cal O}(\{p\}) \Psi & = &
s_{\theta} c_{\theta} (\bar{N_1} \gamma_5 {\cal O}(\{p\})
N_1 + \bar{N_2} \gamma_5 {\cal O}(\{p\}) N_2) \\ & & - i
\frac{1}{2} (s_{\theta}^2- c_{\theta}^2) (\bar{N_1} {\cal
O}(\{p\}) N_2 + \bar{N_2} {\cal O}(\{p\}) N_1) \\ & & \\
          \bar{\Psi} {\cal O}_{\mu}(\{p\}) \Psi& = &
\frac{i}{2} (\bar{N_1} {\cal O}_{\mu}(\{p\}) N_2 - \bar{N_2}
{\cal O}_{\mu}(\{p\}) N_1) \\ & & \\ \bar{\Psi} \gamma_5
{\cal O}_{\mu}(\{p\}) \Psi) & = & \frac{i}{2} (\bar{N_1}
\gamma_5 {\cal O}_{\mu}(\{p\}) N_2 - \bar{N_2} \gamma_5
{\cal O}_{\mu}(\{p\}) N_1 \\ & & \\
          \bar{\Psi} {\cal O}_{\mu \nu}(\{p\}) \Psi& = &
s_{\theta} c_{\theta} ( \bar{N_1} {\cal O}_{\mu \nu}(\{p\})
N_1 + \bar{N_2} {\cal O}_{\mu \nu}(\{p\}) N_2) \\ & & - i
\frac{1}{2}(s_{\theta}^2-c_{\theta}^2) (\bar{N_1} \gamma_5
{\cal O}_{\mu \nu}(\{p\}) N_2 + \bar{N_2} \gamma_5 {\cal
O}_{\mu \nu}(\{p\}) N_1) \\ & & \\
          \bar{\Psi} \gamma_5 {\cal O}_{\mu \nu}(\{p\})
\Psi& = & s_{\theta} c_{\theta} (\bar{N_1} \gamma_5 {\cal
O}_{\mu \nu}(\{p\}) N_1 + \bar{N_2} \gamma_5 {\cal O}_{\mu
\nu}(\{p\}) N_2)\\ & & - i
\frac{1}{2}(s_{\theta}^2-c_{\theta}^2) (\bar{N_1} {\cal
O}_{\mu \nu}(\{p\}) N_2 + \bar{N_2} {\cal O}_{\mu
\nu}(\{p\}) N_1) \end{array} \eeq

     \noindent where ${\cal O}(\{p\})$, ${\cal
O}_{\mu}(\{p\})$ and ${\cal O}_{\mu \nu}(\{p\})$ are general
operators which depend on the momenta $\{p\}$ entering the
vertex and where we have used the self conjugacy properties
of the Majorana fields \cite{gates} .  Since the Majorana
techni-neutrino mass is being treated as a perturbation to
the TC symmetry breaking pattern Eq(\ref{eq:symb}) the
vertices are functions of the dynamical Dirac self energy,
$\Sigma(k^2)$, only.

     The two pion-two vector gauge boson scattering
amplitude and hence $L_{10}$ is again given by the diagrams
in Fig 1 but with all possible positionings of the neutrino
mass eigenstates in the fermion loops.  Remembering that
both the Wick contractions $< \bar{\Psi}_M \Psi_M >$ and $<
\Psi_M \Psi_M>$ are non zero we obtain, for example, the
contributions in Fig 2. We extract the generalized
coefficient, $C$, of the $q^2 g^{\mu \nu}$ term in the
Taylor expansion of the scattering amplitude. When Taylor
expanding the amplitude we assume that $c_{\theta}$ and
$s_{\theta}$ vary slowly with respect to $q^2$ and set all
there derivatives to zero. $L_{10}$ is then given in terms
of $C$ by Eq(\ref{eq:l10=s}). Our full expression for
$L_{10}$ is too long to be reproduced here. However, as a
check we have confirmed the results of Ref\cite{tern} in the
limit $M \rightarrow 0$ ($\theta \rightarrow \pi/4$)
\vspace{3cm}

     \section{Numerical Results}

     In this section we present the numerical results
obtained from the expressions derived for $S$ for a
techni-lepton doublet in Sections 2 and 3. We must estimate
the form of the Schwinger Dyson equations' solution for
$\Sigma(p^2)$ in TC. Following Holdom et al.
\cite{holdom,tern,holdomQCD} we choose a function that
behaves as $1/k^2$ for large momenta and is well behaved as
$k^2 \rightarrow 0$

     \beq \Sigma(k^2) = \frac{(A+1)m^3}{k^2+Am^2}
\label{eq:ansatz} \eeq

     \noindent where A and m are arbitrary coefficients .The
function is normalized so that $\Sigma(m) = m$.  The form of
this solution is plotted in Fig 3 for $A = 1,2,3$ and $A
\rightarrow \infty$.  $\Sigma(k^2)$ corresponds to a hard
mass $m$ when $A\rightarrow \infty$ and comparison to
experimental data \cite{holdomQCD} suggests that $2 \leq A
\leq 3$ gives a good approximation to the solution of the
QCD Schwinger Dyson equations. Values of $A > 3$ give
solutions that fall off more slowly than the QCD solution as
is expected in ``walking" TC theories.

     We first quantify the accuracy of our approximations by
comparing the results from our expression for $S$ when $A
\rightarrow \infty$ with the perturbative one loop
calculation \cite{gatesandtern}. Fig 4 shows this comparison
for a techni-lepton doublet with Dirac mass $m_D$ and
techni-neutrino Majorana mass $M$ (the x axis shows the
normalized Majorana mass $M/m_D$). The Non Local Chiral
Model's prediction for $S$ shows good agreement with the
perturbative result until $M/m_D \sim 2 - 3$. As expected
perturbing the isospin preserving chiral symmetry breaking
pattern is a good approximation when the symmetry breaking
mass terms are $\leq {\cal O}(m_D)$. In TC $m_D \sim {\cal
O}(\Lambda_{TC})$ and hence we expect our results for $S$ to
be good estimates when the Majorana mass $\leq {\cal
O}(\Lambda_{TC})$.  Above $M/m_D \sim {\cal O}(1)$ the
perturbative results diverge as $\log\left( \frac{M_1}{m_D}
\right)$ (where $M_1$ is the mass of the lightest Majorana
neutrino mass eigenstate) as a result of the divergence of
the perturbative diagrams.  The diagrams contributing to $S$
in the Non Local Chiral Model (Fig 1) converge as $M/m_D
\rightarrow \infty$ and hence this behaviour is not
reproduced. Note that the $\log\left( \frac{M_1}{m_D}
\right)$ term only becomes large when the Chiral Symmetry
approximation has broken down.

     In Fig 5 we present the results for the contribution to
$S$ from a techni-lepton doublet with dynamical Dirac self
energy and hard Majorana neutrino mass in the Non Local
Chiral Model for various choices of $A$ in the ansatz for
$\Sigma(k^2)$. If $1 \leq A \leq 3$ is a good approximation
to TC dynamics then in the limit $M = 0$ we find for an
$SU(2)_L$ doublet

     \beq 0.07 \leq S \leq 0.1 \eeq

     \noindent in agreement with the approximation in
Eq(\ref{eq:naives}). It is commonly assumed in the
literature that TC dynamics have the effect of enhancing
perturbative estimates for $S$ by a factor of $2$. However,
our results suggest that the TC dynamics provide a positive
shift to the perturbative results, $0.02 \leq \Delta S \leq
0.05$.

     If the Majorana mass is dynamically generated then it
too will have momentum dependence of the form in
Eq(\ref{eq:ansatz}). Fig 6 shows the deviations from the
estimates with a hard Majorana mass in the results for $S$
due to this momentum dependence (we show results for $A=3$
as an example). The new estimate tends to the result with a
hard Majorana mass for low momenta since the momentum
dependence is negligible in comparison to the large Dirac
self energy. For large Majorana masses the Dirac self energy
falls off whilst the Majorana mass is still flat and again
the results tend to the hard limit. The deviation in $S$
from the hard limit are at most $\sim 0.01$.

     Finally we consider the consequences of these results
in one family TC models \cite{TC,kingtype}. Recent work
\cite{appelandtern} has shown that realistic values for $S$
and $T$ can be obtained from mass splittings in the
techni-lepton sector and from Goldstone boson contributions.
In models in which the techni-quarks and leptons interact
equally with the TC gauge bosons \cite{TC} the techni-quark
self energy can be enhanced by QCD interactions by a factor
of $2 - 5$ relative to the techni-lepton self energy
\cite{sigq}. The $W$ and $Z$ masses are then generated
almost exclusively by the techni-quark condensates since,
for example

     \beq M_Z^2 = \frac{g^2+g'^2}{4} \left( \frac{1}{2}f_N^2
+\frac{1}{2}f_E^2 + 3f_Q^2 \right) \eeq

     \noindent where $f_N$,$f_E$ and $f_Q$ are the Goldstone
boson decay constants associated with the Goldstone bosons
with constituent techni-neutrinos, techni-electrons and
techni-quarks respectively.  Large mass splittings in the
techni-lepton sector can, therefore, give negative
contributions to $S$ without contributing a large positive
contribution to $\delta \rho_*$.

     We consider the contributions to $S$ and $\delta
\rho_*$ from techni-neutrino Majorana masses, M, in such a
scenario in Table 1. We extrapolate our results for $S$ when
$M \leq {\cal O}(\Lambda_{TC})$ to larger values of M by
shifting the perturbative results by
$[S_{NLCM}(M=0)-S_{pert}(M=0)]$ as suggested by the results
in Fig 3.  The contributions to $\delta \rho_*$ are
estimated using the Dynamical Perturbation Theory \cite{DPT}
results from Ref\cite{majT}.  The techni-quarks are assumed
to have degenerate Dirac self energies as are the
techni-leptons. The parameters $m_L$, $m_Q$ and $M$ in the
ansatz for $\Sigma(k^2)$ are scaled in Table 1 by the mass
scale $\Lambda$ which must be tuned in a particular TC model
to give the experimentally measured $Z$ mass. Three
estimates for $S$ and $T$ are given, the naive results from
Eq(\ref{eq:naives}) and the Non Local Chiral Model results
with $A=2$ and $A=3$ in the ansatz for $\Sigma(k^2)$.

     The techni-family's contribution to $S$ can be reduced
by a factor of $\sim 4$ without violating the experimental
upper bound on $\delta \rho_*$ by including the Majorana
mass for the techni-neutrino. The largest negative
contributions to $S$ are achieved at the expense of a large
Majorana mass ($M/m_D \sim 10$) in which case the lightest
Majorana mass eigenstate may fall below the current
experimental lower bound in some models.  However,
additional negative contributions to $S$ from splitting
between the Dirac self energies in the lepton doublet and
from Goldstone boson contributions \cite{appelandtern} are
not included in these estimates. If these additional
negative contributions to $S$ were included in a realistic
TC model a more reasonable value for $M/m_D$ ($\sim 5$)
could be chosen whilst maintaining physical values for $S$
and $T$

     \begin{tabular}{|c|c|c|c|c|c|c|c|c|} \hline $m_L/
\Lambda$ & $m_Q/ \Lambda$ & $M/ \Lambda$& $S_{naive}$ &
$\delta \rho_{* naive} \%$ & $S_{A=2}$ & $\delta \rho_{*
A=2} \%$ &$S_{A=3}$ & $\delta \rho_{* A=3} \%$ \\ \hline & &
& & & & & & \\ $1$ & $1$ & $5$ & $0.4 N_{TC}$ & $0$ & $0.23
N_{TC}$ & $0.69$ & $0.19 N_{TC}$ & $0.81$ \\ $1$ & $1$ &
$10$ & $0.4 N_{TC}$ & $0$ & $0.16 N_{TC}$ & $2.43$ & $0.12
N_{TC}$ & $2.55$ \\ & & & & & & & & \\ $1$ & $2$ & $5$ &
$0.4 N_{TC}$ & $0$ & $0.23 N_{TC}$ & $0.26$ & $0.19 N_{TC}$
& $0.30$ \\ $1$ & $2$ & $10$ & $0.4 N_{TC}$ & $0$ & $0.16
N_{TC}$ & $0.76$ & $0.12 N_{TC}$ & $0.77$ \\ & & & & & & & &
\\ $1$ & $3$ & $5$ & $0.4 N_{TC}$ & $0$ & $0.23 N_{TC}$
&$0.16$ & $ 0.19 N_{TC}$ & $0.16$ \\ $1$ & $3$ & $10$ & $0.4
N_{TC}$ & $0$ & $ 0.16 N_{TC}$ & $0.34$ & $0.12 N_{TC}$ &
$0.33$ \\ & & & & & & & & \\ \hline \end{tabular}

\begin{center} Table 1. Values for $S$ and $\delta \rho_*$
for various techni-fermion
mass spectra in one family TC models. \end{center}

     \section{Conclusions}

     In this paper we have developed non perturbative
estimates for the electroweak radiative correction parameter
$S$ in dynamical symmetry breaking models with Majorana
neutrino masses. The Majorana mass terms were treated as
perturbations to the isospin preserving chiral symmetry
breaking pattern in Eq(\ref{eq:symb}) of a Non Local Chiral
Model of the strong dynamics that break electroweak
symmetry.  Comparison with the perturbative one loop results
for $S$ with hard mass terms suggest that the approximations
made hold when the Majorana mass is below the chiral
symmetry breaking scale. Estimates for $S$ in Technicolour
theories were then made with a suitable ansatz for the
techni-fermion self energies. The results suggest that the
TC dynamics have the effect of shifting the perturbative
results for $S$ by a factor of $0.02 - 0.05$ per doublet.
In walking TC theories the shift would be lower depending on
the precise fall off of the techni-fermion self energies. We
conclude that Majorana neutrino masses can give negative
contributions to the $S$ parameter in non-perturbative
theories.

     Techni-fermion mass spectra were proposed in one family
TC models in which the negative contributions to $S$ from
Majorana neutrino masses reduced the estimate for $S$ for
the full family by a factor of $\sim 4$ without giving rise
to an unphysically large contribution to $\delta \rho_*$.
These negative contributions to $S$ are achieved with
natural choices of mass scales and without fine tuning. We
do require though that the techni-quarks dominate
electroweak symmetry breaking as in models such as those in
Ref\cite{TC}. Extended TC models such as those in
Ref\cite{kingtype} in which the large top bottom mass
splitting is generated by a large techni-up techni-down mass
splitting require that the techni-lepton sector dominates
electroweak symmetry breaking. These models will be more
tightly constrained by the $\delta \rho_*$ estimates.
Estimates of $S$ and $T$ in TC models accounting for mass
splittings in the techni-lepton sector, from Goldstone boson
contributions \cite{appelandtern}, and now these results for
techni-neutrino Majorana masses suggest that parameter
ranges exist in TC models with realistic fermion mass
spectrums that give realistic values of $S$ and $T$.
\newpage

     \noindent {\bf Acknowledgments}

     The authors would like to thank Steve King for helpful
discussions.  One of us (NJE) thanks SERC for financial
support.

\end{document}